\documentstyle[12pt]{article}

\begin{document}

\baselineskip = 18pt plus 1pt minus 1pt

\title{Quark Model Form Factors for Heavy Quark Effective Theory}

\author{ FE Close\\
\makebox[5cm]{}
\\
{\em Rutherford Appleton Laboratory,}\\
{\em Chilton Didcot, Oxon OX11 0QX, England.}
\\
\\
          A Wambach
\\
\makebox[5cm]{}
\\
{\em Theoretical Physics,}\\
{\em Department of Physics,} \\
{\em 1 Keble Road, OXFORD, OX1 3NP, England.} \\
\\
\\
}
\vspace{2cm}
\date{RAL--93--022\\OUTP 93 06 P\\ 28.04.1993}

\maketitle

\begin{abstract}
\noindent We show how both the spectroscopy of heavy-light hadrons and the
slope of the Isgur-Wise function can be simultaneously fitted when
Heavy Quark Effective Theory is matched onto dynamical quark
models, including careful treatment of Wigner rotations. Working consistently
to order $\vec{v}^2$ and using the parameters from hadron spectroscopy
as input, we determine the slope parameter $\rho $ for s-wave
transitions to be $ 1.19 \pm 0.02 $. This agrees with the empirical value
for $b \rightarrow c$ transitions.
\end{abstract}
\pagestyle{empty}

\newpage
\pagenumbering{arabic}
\pagestyle{plain}

\section{INTRODUCTION}

In recent years considerable experimental and theoretical effort
has been  invested  to understand the physics of hadrons
containing a heavy quark. Applications include the
determination of important physical quantities such as elements of the
Kobayashi-Maskawa matrix. A
fundamental problem for theory is to extract data at
quark level from experiments that involve hadrons.

Heavy Quark Effective Theory (HQET) \cite{isgur1} has simplified the
analysis  by showing that in the limit
$M_Q \rightarrow \infty$ the spins of the heavy and light degrees of freedom
decouple and that a single form factor (the ``Isgur--Wise function'') describes
weak decays of heavy hadrons. In particular this promises that ratios of
hadron decays {$\Gamma(H_1\rightarrow H_2l\nu)/\Gamma(H_3 \rightarrow
H_2l\nu)$}
 may
provide direct measures of the corresponding CKM quark matrix elements
$V_{12}/V_{32}$. However, in practice the $M_Q\rightarrow\infty$ approximation
is bad for strange quarks certainly and arguable for charm. In order to make
direct connection between heavy hadron and the corresponding quark amplitudes
we need knowledge of the Isgur--Wise function $\xi(y)$ (where $y=v \cdot v'$
and
$v_{\mu}^{(')}$ is the four--velocity of the heavy meson before (after) the
transition), and/or insights into $1/M_Q$ corrections to HQET.

HQET does not predict the $\xi(y)$. The form factor is normalized to unity at
the zero recoil point where the meson retains its velocity during the
transition.
For small, non zero, recoil it is conventional to write
\begin{equation}
\xi(y) = 1 - \rho^2(y-1) + 0((y-1)^2)
\end{equation}
where $\rho$ is the ``slope parameter'' or ``charge radius''.

There have been several attempts to calculate $\rho$ from theory and models
 \cite{voloshin,ali,neubert2},
but all appear to underestimate its magnitude compared to the empirical value.
A recent calculation of $\xi(y)$ on the lattice appears to be consistent with
data, albeit with large uncertainties at present \cite{ukqcd}; QCD sum rules
have also been applied with rather different conclusions
\cite{blok,rad,neubert3}.

In particular, Isgur et al (ISGW, ref[\citescora}]) have discussed the
spectroscopy
of $Q\bar{q}$ within a quark model that has proven remarkably robust in
describing mesons \cite{godfrey}. However, when the form factor is computed in
this model it appears to fail empirically. A central result in the
present  paper
is that some
previous estimates of $\rho$ in the quark model underestimated it due to
an inconsistent non--relativistic approximation.
Specifically
we shall show that the quark model, when fitted to heavy meson spectroscopy
(as in \cite{scora}) yields a value of $\rho$ that agrees with form factor data
when the operators for current transitions for {\it{ composite hadrons with
recoil}} are carefully expanded to $0(\vec{v}^2)$.

We begin by reviewing the data and then the existing quark model calculations.
The main thrust of the paper is to show how the apparent failure in the latter
case is due to an incomplete treatment and to demonstrate how the quark
model can be consistently applied for heavy--light hadrons.

In refs. \cite{neubert1,rosner,mannel}
data on $B \rightarrow Dl\nu$ have been fitted
over a limited range of $y \not=1$
with the assumption that they approximate the IW function ($m_Q \rightarrow
\inf$). These authors make various choices for
the functional form of its $y$ dependence. These differ at $y>>1$ but agree
within the errors over the existing range, $y \approx 1$. however, the
resulting
fits lead to different coefficients of the charge radius, eq(1). This is not
immediately apparent as these references use different conventions and so
we rewrite their parametrisations so that they have
a common definition of $\rho$ as in the leading order expansion at eq(1). The
different ansatze produce correspondingly slightly different values for $\rho$
with the following results:
\begin{eqnarray}
{\rm Neubert \; \cite{neubert1}} & \xi(y) = e^{-\rho^2(y-1)} & \rightarrow \rho
= 1.07 \pm 0.22\\
\nonumber\\
{\rm \cite{neubert1,neubert2}}& \xi(y) =
\frac{2}{y+1}e^{-(2\rho^2-1)\frac{y-1}{y+1}} & \rightarrow \rho = 1.14 \pm
0.23\\
\nonumber\\
{\rm (pole \; ansatz) \; \cite{neubert1}} & \xi(y) = (\frac{2}{y+1})^{2\rho^2}
& \rightarrow \rho = 1.19 \pm 0.25\\
\nonumber\\
{\rm Rosner \; \cite{rosner}} & \xi(y) = \frac{1}{1+\rho^2(y-1)}
& \rightarrow \rho = 1.29 \pm 0.28\\
\nonumber\\
{\rm Mannel \; \cite{mannel}} & \xi(v) = 1 -
\frac{\rho^2}{2}(y-1)(y+1) & \rightarrow \rho = 0.99 \pm 0.04
\end{eqnarray}

Neubert\cite{neubert1} also makes a linear fit to the data:

\begin{equation}
 \xi(v) = 1 - \rho^2(y-1) \rightarrow \rho = 0.92 \pm 0.21
\end{equation}
 On general grounds the universal form factor is
expected to have positive curvature for all $y > 1$ and so $\rho$ obtained
from the linear fit is expected to be a very conservative {\it lower limit}
\cite{neubert1}. Present data cannot distinguish among these hypothesised
functional
forms. We are here concerned only with the coefficient of the $(y-1)$ term and
the range of values in eqs (2-7) may be taken as a guide to the systematic
uncertainty in this quantity. This has important implications for the
non-relativistic
quark model (NRQM) as we now show.

In the NRQM where it is assumed that
the initial meson is at rest and after the
transition moves with the velocity $v'=P/M$,
 the Isgur-Wise function describes the overlap between two light
quarks in an s-wave state, where one of the quarks is moving relative
 to the other. It takes the form:

\begin{equation}
\xi(v) = \int \phi^*(k') \phi(k'+mv') {\rm d}^3k' = \int
\phi^*(x)\phi(x)e^{-im\vec{v'}\vec{x}} {\rm d}^3x \\
\end{equation}
\\
In the NRQM this comes about because
\begin{equation}
\phi(p)=\int{\rm d}^3r_1{\rm d}^3r_2 e^{i\vec{p}_1\vec{r_1}}
e^{i\vec{p}_2\vec{r_2}}\phi(\vec{r}_1-\vec{r}_2)  =
\int{\rm d}^3R{\rm d}^3r e^{i\vec{P}\vec{R}}e^{i\vec{p}\vec{r}}\phi(r)
\end{equation}
 that is, the overall momentum $P=p_
1+p_2$ separates itself from the internal momentum
$p=\frac{1}{m_1+m_2}(m_2p_1-m_1p_2)$.

In the ISGW model \cite{scora} a Gaussian ansatz
$\phi(r)=exp(-\frac{\beta^2r^2}
{2})$ is
used to fit the heavy quark spectroscopy. $\beta$ is related to the excitation
energy in an harmonic oscillator spectrum. The fit of ref.\cite{scora} assumes
a linear plus
Coulomb potential, and their wavefunctions are treated as a superposition of
Gaussians. They require that, if the mass of the degree of freedom represented
 by the light quark is 330MeV, then $\beta=0.34$GeV for
 K systems and
$\beta =0.4$GeV for B and D hadrons of interest here (which are essentially
the values for the $m_Q \rightarrow \inf$ limit; note that refs 5-10 implicitly
or explicitly assume that the B system equates with this limit).

If this wavefunction is used in eq(8) then
\begin{equation}
\xi(v) = exp( -\frac{\vec{p}^2}{4\beta^2}) \equiv exp(-
\frac{m^2}{2\beta^2}(y-1)) =
         1 - \frac{m^2}{2 \beta^2} (y-1) + O((y-1)^2)
\end{equation}
This is equal to eq(1) if $\rho^2 = \frac{m^2}{2\beta^2}$. Inserting ISGW
values for $m$, $\beta$ yields $\rho = 0.57$, significantly below the
experimental value extracted from a Gaussian function (ref\cite{neubert2},
eq(2)).Even if one only accepts the result up to the leading term
in $(y-1)$ the value of $\rho$ is too small independent of parametrisation
chosen in eqs(2)--(7).

Ref.\cite{scora} anticipated that their model would be incomplete for large
recoil
(such as in $b \rightarrow u$) and modified the computed slopes of all form
factors by a universal multiplicative factor chosen by fitting the observed
pion form factor, on the grounds that all heavy-light hadrons have similar
size. The effect is that $\rho$ is renormalised \cite{scora}
$\rho \rightarrow \frac{\rho}{0.7} \approx 0.8$

Even after this ad hoc adjustment the value is only at the lower edge of the
value as extracted empirically (refs \cite{neubert1,neubert2,rosner,mannel},
eqs. (2)--(6) or even eq.(7)).
 As some of the transitions
that are input to the extraction of CKM matrix elements involve non zero
recoil it seems desirable to examine this question further.

Ref\cite{li} delineated the kinematic range over which a non-relativistic
description of hadron transitions may be consistent, in the sense of preserving
the necessary low energy theorems. In ref\cite{close} they showed how these
constraints have implications for the matching of heavy quark effective theory
 and quark model descriptions when applied to current-hadron interactions at
$O(\vec{v^2})$. They referred to this matching as ``Effective Heavy Quark
Theory" or EHQT.
As we shall see in this paper, when the EHQT ideas of ref\cite{close} are
applied to the present problem,
including the necessary separation of relative and overall momenta
variables and consequent Wigner--rotation \cite{close}, the value of $\rho$
increases significantly.
 The step from NRQM
to relativistic kinematics increases $\rho$ to about 1
(as has already noted by Neubert \cite{neubert1} following \cite{wirbel},
 but which does not seem to have been widely appreciated).
The approach of ``Effective Heavy Quark Theory'' -- when explicit quark
model states are boosted consistently so as to match with Heavy Quark
Effective Theory, shows how this relativistic kinematic effect arises but also
highlights that Wigner--rotation
modifies this further, leading to $\rho = 1.19 \pm 0.02$ .

The set--up of this paper is as follows. We first briefly repeat the
description of meson states in HQET. Then we explain how the Wigner--rotation
of the light quark appears in  usual quark model descriptions. The next step
is to match these two theories and derive a  formal representation of heavy
mesons from which we can derive the Isgur-Wise function.

\section{MATCHING QUARK MODELS AND HQET}

Ref.\cite{close} showed how to match HQET onto explicit quark model
descriptions
of current induced transitions.
We shall first review this matching
in a form that makes contact with the widely used HQET
formalism.

 An essential assumption is that in the
infinite mass limit the heavy quark does not undergo spin interactions
with the light degrees of freedom \cite{isgur1}. Interaction with an external
field $W_{\mu}$ involves a
current $J_{\mu}$ which has the form:
\begin{equation}
\begin{array}{ccc}
<D(v')|\overline{Q'(v')} \Gamma_{\mu} Q(v) |B(v)>& =&
\overline{Q'(v')}\Gamma_{\mu}Q(v)\times \overline{q'(v')}T(v,v')q(v)\\
\\
& =& Tr [ \overline{M'(v')}\Gamma_{\mu} M(v) T(v,v')]
\end{array}
\label{transition}
\end{equation}
where $M(v)= Q(v)\overline{q(v)}$ and $Q(q)$ is the spinor for
the heavy (light) quark or antiquark. The coupling for the light quark
sector $(T)$ contracts with the mesons' spin--wave functions to give
a scalar form factor, depending only on $v \cdot v'$ (see eg.\cite{balk,falk}).

Mesons in HQET are therefore described in matrix form. In the rest frame,
the matrix description  of s--wave mesons is written \cite{falk,bjorken}:
\begin{equation}
M(v=(1,\vec{0})) = \left( \begin{array}{cc} 0 & X\\
                                            0 & 0 \end{array} \right)
\\
\label{mesonmatrix1}
\end{equation}
where $X=-1$ for the pseudoscalar state $(0^{-+})$ and $X=\vec{\sigma} \cdot
\vec{\epsilon}$ for the vector state $(1^{--})$. When boosted to velocity
$v_\mu=(v_0,\vec{v})$ one finds for the s-wave mesons:
\begin{equation}
\begin{array}{ccc}

M_0(v) & = & \frac{1}{\sqrt{2}} \frac{1}{2}(1+\not{\!v})\gamma_5 \\
\\
M_1(v) & = & \frac{1}{\sqrt{2}} \frac{1}{2}(1+\not{\!v})\not{\!\epsilon}
\end{array}
\label{mesonmatrix2}
\end{equation}
 The generalisation to $L \neq 0$ is immediate \cite{falk,balk}.

This is standard. New insights follow when we match this to the quark model
by following the prescription of EHQT \cite{close}.

For meson
transitions it is traditional to use equation (\ref{transition}) as
the starting point. But as discussed in refs \cite{close,brodsky}  the spinor
of the light antiquark has to be defined with
care. An example of this can be seen in the work of Golowich
 et al. \cite{iddir}.
These authors noted  that in certain circumstances the neglect
of Wigner-rotations associated with recoil in current induced transitions
can cause explicit inconsistencies and contradictions in
the calculation of form factors for
meson decays, in particular $B \rightarrow D^*l\nu$ as here.
An advantage of the Effective Heavy Quark Theory approach is that such effects
are incorporated from the outset and throughout the calculation. We now
illustrate this.

If we go to the zero--binding limit, the spinor for the light antiquark
in a heavy meson at rest  takes the form:
\begin{equation}
v(p_{q},s_{q})  =  \sqrt{\frac{\omega_{q} + m_{q}}{2m_q}} \left(
 \begin{array}{c}
          \frac{\vec{\sigma}\cdot\vec{k}}{\omega_{q} + m_{q}}\phi_{q}\\
                                           \phi_{q}
\end{array}\right)
\label{spinor}
\end{equation}
where $\omega_q = \sqrt{m_q^2+\vec{k}^2}$ and $m_q$ is the mass of the
light quark.
Now boosting the meson and therefore also the light quark
 to the velocity $v$ the spinor transforms into
\begin{equation}
v(p_{q},s_{q}) =\sqrt{\frac{1+v_{0}}{2}} \sqrt{\frac{\omega_{q} + m_{q}}{2m_q}}
                \left( \begin{array}{c}
 (\frac{\vec{\sigma} \cdot \vec{v}}{1+v_{0}}+\frac{\vec{\sigma} \cdot \vec{k}}
                                 {\omega_{q} + m_{q}})\phi_{q}\\
 (1+\frac{\vec{\sigma} \cdot \vec{v}\vec{\sigma} \cdot \vec{k}}{(1+v_{0})
                                 (\omega_{q}+m_{q})}) \phi_{q}
\end{array}\right)
\label{spinor1}
\end{equation}

 The underlying reason for performing these two boosts (i.e. the $k$ boost
followed by the $v$ boost) separately as distinct from directly
boosting the light quark to its full four-momentum $\tilde{k}$ is that
the rest frame of the meson is the frame where the spins
of the heavy and light quark are defined. Therefore one has to set up
the description of the meson in this frame, identify the
appropriate states in this frame and
 then boost the whole system to its final velocity
$v$ \cite{close}.

 As is obvious from equation (\ref{spinor1}), if one overlooks the first boost,
one fails consistently to account for the Wigner--rotation of the light quark.
A similar boost arises for the heavy quark, but at order
$1/M$. This effect for the {\it heavy} quark does therefore not appear
in the infinite mass limit but has to be included for first order
corrections which can be significant, especially when applied to strange quarks
in the final state.

To make the matching of HQET and the quark model (thus ``EHQT")
we return to
equation (\ref{mesonmatrix1}). With the matrix description $ M=Q\bar{q} $  (or
$M=u\bar{v}$ in the quark model
language) it is possible to include the light quark boost directly. This then
takes the
form:
\begin{equation}
M(v=(1,\vec{0}))  =  \left( \begin{array}{lr} 0 & X\\
                                             0 & 0  \end{array} \right)
 \left( \begin{array}{lr} 1 & -\frac{\vec{\sigma} \cdot
\vec{k}}{m_q+\omega_q}\\
 -\frac{\vec{\sigma} \cdot \vec{k}}{m_q+\omega_q} & 1 \end{array} \right)
\sqrt{\frac{m_q+\omega_q}{2m_q}}
\label{meson1}
\end{equation}
If we scale the light quark momentum and energy by its mass $(m_q)$, then
since only the upper right corner in the first matrix is
non-zero we can decompose $M$ into an alternative form,
\begin{equation}
M = \left( \begin{array}{lr} 0&X\\
                             0&0 \end{array} \right)
(1-\not{\!k}) [2(1+\omega)]^{-\frac{1}{2}}
\end{equation}
Note that we still use $k$ and $\omega$ for the scaled
momentum and energy of the light quark to avoid further indices.
Using this form and boosting the whole system  we derive
explicit forms for the S wave states,

\begin{equation}
\begin{array}{ccc}
0^{-+} & =  & \frac{1}{\sqrt{2}} \frac{1}{2}(1+\not{\!v})\gamma_5
  (1-\not{\!\tilde{k}}) [2(1+\tilde{k}.v)]^{-\frac{1}{2}}\\
\\
1^{--} & = & \frac{1}{\sqrt{2}} \frac{1}{2}(1+\not{\!v})\not{\!\epsilon}
(1-\not{\!\tilde{k}}) [2(1+\tilde{k}.v)]^{-\frac{1}{2}}\\
\end{array}
\label{meson2}
\end{equation}
\\
Comparing with eqs(\ref{mesonmatrix2})  we see
that they are exactly the
same meson states  but with an additional factor
$(1-\not{\!\tilde{k}})[2(\tilde{k}. v+1)]^{-\frac{1}{2}}$ to the right, where
$\tilde{k}_{\mu}$ is
the boosted four momentum of the light quark (light degrees of
freedom) scaled by its mass (here we used the
identity $\tilde{k}.v=\omega$). These ideas can be  generalised to
p-waves where again we find the same additional factors \cite{achim}
arising from the (boosted) relative momenta.

Eq(\ref{meson2}) is essentially
 a {\it covariant} expression
for the meson which makes the
Wigner rotation of the light quark explicit. It is this form, with explicit
appearance of $\not{\!k}$ to the right of $\not{\!v}$, that encodes the
EHQT in the sense of \cite{close}. The
possibility of correlations between the light quark and the heavy
quark sector is automatically accomodated if we write mesons in this form;
contrast this with many calculations in quark models where one deals
with the light and the heavy quark separately. This explicit exhibition of
such spin correlations
will be useful in the next sections when the form factor is evaluated.

In the complete description of the meson wavefunction, the internal
momentum distribution $\phi(\tilde{k})$ is included and is normalised
such that $\int {\rm d}^3\tilde{k} |\phi(\tilde{k})|^2 = 1$. With $\phi$ being
a scalar, and using $\frac{{\rm d}^3\tilde{k}}{\tilde{\omega}} =
\frac{{\rm d}^3k}{\omega}$, one can easily convince oneself that this
wavefunction has to be multiplied by
$(\frac{\tilde{k}.v}{\tilde{\omega}})^{\frac{1}{2}}$ to satisfy this
normalization condition\footnote{We would like to thank A. Le Yaouanc and
J.--C. Raynal for pointing this out - see also ref\cite{pene}.}.

\section{ISGUR--WISE FUNCTION }

We shall now illustrate how these quark model matrices are applied
by calculating the Isgur--Wise function
to the order $\vec{v'}^2$ (where $\vec{v'}$ is the velocity of the
meson after the transition in the rest frame of the initial meson).

First we make some remarks about the kinematics of the transition.

In the most general case the meson has the four-velocity $v_{\mu}$ before and
$v_{\mu}'$ after the transition while the light degrees of freedom
(which are in the quark model approach synonymous to the light quark) have
the four-momentum $\tilde{k}_{\mu}$ before and $\tilde{k}'_{\mu}$
after the transition (where we assume throughout that
these momenta are scaled by the mass of the light quark).
We defined the light quark spinor
by starting with an internal momentum $\vec{k}$ and than
boosting to the meson velocity. Therefore the usual quark model assumption that
the four
momenta of the light degrees of freedom are conserved during the
transition implies that $\tilde{k}_{\mu} = \tilde{k}'_{\mu}$, and hence that

\begin{equation}
\begin{array}{rcl}
\omega v_0 + \vec{v}\vec{k} &=& \omega' v'_0 + \vec{v}'\vec{k}'\\
\\
\omega \vec{v} + \vec{k} + \frac{\vec{k}\vec{v}}{1+v_0} \vec{v} & = &
\omega' \vec{v}' + \vec{k}' + \frac{\vec{k'}\vec{v'}}{1+v_o'} \vec{v}'
\\
\end{array}
\label{kinematic1}
\end{equation}
In the rest frame of the decaying meson  (i.e. $v_{\mu}=(1,\vec{0})$,
$v_{\mu}=(v'_0,\vec{v}')$) one gets the kinematic constraints:

\begin{equation}
\begin{array}{ccl}
\omega & = & \omega' v'_o + \vec{k}'\vec{v}'\\
\\
\vec{k}  & = & \omega' \vec{v}' + \vec{k}' + \frac{\vec{k'}\vec{v'}}{1+v_o'}
\vec{v}'\\
\\
\end{array}
\end{equation}

or equivalently

\begin{equation}
\begin{array}{ccl}
\\
\omega' & = & \omega v'_0 - \vec{k}\vec{v}'\\
\\
\vec{k}'  & = & -\omega \vec{v}' + \vec{k} + \frac{\vec{k}\vec{v'}}{1+v_o'}
\vec{v}'\\
\\
\end{array}
\label{kinematic2}
\end{equation}

At $O(\vec{v}^2)$ these already take one beyond the standard non-relativistic
formalism and hence
these relativistic transformations will modify the calculation of
$\rho$ at eq(8). If one uses
$\vec{k} \rightarrow \vec{k'} + (\frac{\vec{v'}\vec{k'}}{1+v'_0} + \omega')
 \vec{v'}$ instead of $\vec{k} \rightarrow \vec{k'} + m_q\vec{v'}$
in eq(8) the
 slope parameter turns out to be $\rho = 1.03$ , an enhancement
 compared with (3) of about 100\%. This has already been noted in
\cite{neubert2}.

We can understand this result heuristically as follows. In the non-relativistic
case we found $\rho^2 = \frac{m^2}{2\beta^2}$ and this becomes
$\rho^2 \simeq \frac{\omega^2}{2\beta^2}$ relativistically. Since
$\omega^2 \equiv m^2+<\vec{k}^2>$ and $<\vec{k}^2> \approx O(\beta^2)$ then
\begin{center}
$\rho_{rel}^2 \simeq \frac{m^2 + <\vec{k}^2>}{2\beta^2} \approx \rho_{NR}^2
+\frac{1}{2} \approx 1$.
\end{center}
However, Wigner rotations of spin have not been accounted for consistently
up to this point. The matrix
formalism of section 2 shows this explicitly as we shall now demonstrate.
This will increase the predicted value for $\rho$ by a further $O(10\%)$.

Following the approach given above, the s-wave to s-wave transition
 elements in the unified formulation  take the following form:
\begin{equation}
\begin{array}{cc}
&{\rm Tr} [(1-\not{\!\tilde{k'}})L'\frac{1}{2}(1+\not{\!v'}) \Gamma_{\mu}
 \frac{1}{2}(1+\not{\!v})L(1-\not{\!\tilde{k}})]\phi^{*}(\tilde{k'})
\phi(\tilde{k})\\
\\
=&   {\rm Tr}[L'\frac{1}{2}(1+\not{\!v'})\Gamma_{\mu} \frac{1}{2}(1+\not{\!v})
L(1-\not{\!\tilde{k}})2]\phi^{*}(\tilde{k'})\phi(\tilde{k})

\end{array}
\label{transition2}
\end{equation}
where $L(L')$ is either $\gamma_5$ or $\not{\!\epsilon}$ and the kinematical
 factors  are included in
the wave functions $\phi$. For the equality
we used that $\tilde{k}_{\mu} = \tilde{k}'_{\mu}$ and $\tilde{k}.\tilde{k}=1$.

Using this expression, going to the frame defined above with an
additional constraint, namely that $\vec{v'}$ is parallel to the
z-axis (i.e. $\vec{v'}=(0,0,v')$),
inserting the kinematic constraints (\ref{kinematic2}), the light quark sector
contribution to the overlap of initial and final states takes the form:
\begin{equation}
\int {\rm d}^3 k{\rm d}^3 \tilde{k'}
(1-\not{\!k})[(1+\omega v'_o-v'k_z)(1+\omega)]^{-\frac{1}{2}}
(\frac{\omega v_0'-k_z v'}{\omega})^{\frac{1}{2}}
(\beta_1\beta_2\pi)^{-\frac{3}{2}} e^{-\frac{k'^2}{2\beta_1^2}}
e^{-\frac{k^2}{2\beta_2^2}}\delta^3(\tilde{k'}-\vec{k})
\label{isgurwise1}
\end{equation}
where we used explicit s-wave quark model wave functions as in \cite{scora}
and the delta function and normalisation follow the eq(5.8) of
ref\cite{brodsky}.

In the HQET limit $\beta_1 =\beta_2$ but we shall retain
different harmonic oscillator strengths here because in practice for
charm and strange quarks, with non-infinite masses, these strengths differ
considerably (see \cite{scora}). Expanding the exponential and the
inverse square root in $v'$ and keeping only terms proportional
to $v'$ and $v'^2$ one derives

\begin{equation}
\begin{array}{c}
( \pi \beta_1\beta_2)^{-\frac{3}{2}} \int {\rm d}^3 k
[(1-\omega \gamma_0+\vec{k}\vec{\gamma})(1+\omega)^{-1}e^{-Ak^2}
\times\\
\\
(1 + v'k_z(\frac{-1}{2\omega}+\frac{\omega}{\beta_1^{2}}) + v'^2 (
\frac{-\omega}{4(1+\omega)} - \frac{\omega^2}{2\beta_1^2} +
k_z^2(\frac{3}{8(1+\omega)^2} -\frac{1}{8\omega^2} -
\frac{\omega}{2\beta_1^2(1+\omega)}
-\frac{1}{2\beta_1^2}+\frac{\omega^2}{2\beta_1^4})))]
\end{array}
\label{isgurwise2}
\end{equation}
$A$ is defined to be $\frac{1}{2}(\beta_1^{-2}+\beta_2^{-2})$.
Before we complete the calculation, note first the structure of the
expression in (\ref{isgurwise2}). It is obvious that the final result will have
the form:
\begin{equation}
B + C\gamma_0 + v'D\gamma_{z} = B + (C+v'_0D)\not{\!v} - D\not{\!v'}
\end{equation}
where on the RHS we specialised to the frame $v=(1,0,0,0)$ and
$v'=(v'_0,0,0,v')$. Because at the end of the calculation we have to insert
 this expression into (\ref{transition2}) we now can contract $\not{\!v}$ and
$\not{\!v'}$
with  $(1+\not{\!v})$ and $(1+\not{\!v'})$. This has the consequence that the
Isgur--Wise function in this quark model approach takes the form:
\begin{equation}
\xi(v) = B - (C+v'_0 D) + D = B - C - \frac{1}{2} v'^2 D
\end{equation}
This has the effect that
\begin{equation}
1-\omega \gamma_0 + k_{z}\gamma_{z} \rightarrow 1+ \omega - \frac{1}{2}v'k_{z}
\end{equation}
At this stage the relativistic form of a heavy meson in the quark model
developed in (\ref{meson1}),(\ref{meson2}) reduces the calculation for the
light quark sector considerably, because  of the direct contraction with the
heavy quark sector.
Doing the calculation one gets for the form factor:
\begin{equation}
\begin{array}{lcc}
\xi(v') & = & (\frac{\beta_1^2 + \beta_2^2}{2\beta_1\beta_2})^{-\frac{3}{2}} (
1+v'^2[-\frac{1}{2}\beta_1^{-2}- A^{-1}(\beta_1^{-2}-\frac{1}{4}\beta_1^{-4})
 +\frac{5}{8}A^{-2}\beta_1^{-4}])\\
\\
&& +v'^2(\pi\beta_1\beta_2)^{-\frac{3}{2}} \int (\frac{-\omega}{4(1+\omega)}
-\frac{k_z^2}{8\omega^2} + \frac{k_z^2}{8(1+\omega)^2})e^{-Ak^2} {\rm d}^3k\\
\\
\end{array}
\end{equation}
where $A$ is  defined as in (\ref{isgurwise2}). This leads to $\rho = 1.19$
where we fitted the harmonic oscillator strength with the value given by
\cite{scora}.

Relativistic recoil and Wigner rotation of the light
quark have each increased the slope of the Isgur-Wise function. The
non--relativistic quark model gave $\rho = 0.57$; taking account of
relativistic
kinematics for the light quark while ignoring
spin--interaction increased this to $\rho = 1.03$  and including a rotation
of the light quark spin increases $\rho$ to 1.19. This qualitative behaviour
is quite reasonable. The Wigner spin rotation effectively decreases the
probability
for the final state pseudoscalar to overlap with the initial and hence the
form factor falls faster, or equivalently $\rho$ increases,
 than when this effect is ignored.

The only free
parameter in this model is the ratio of the mass of the light
antiquark and the coupling strength $(\frac{m}{\beta})$.
 For $\beta=360$MeV (460MeV) and $m$=330MeV we get
$\rho=1.23 (1.17)$.
 This result
is insensitive to varying this ratio over
a wide range of reasonable values.

In general (\ref{transition2}) allows the
calculation of the form factor to an arbitrary order in $v'$, but the quark
model loses much of its predictive power  when $y >> 1$ because the
assumption $\tilde{k}_\mu=\tilde{k}'_\mu$ is no longer justified. As
pointed out in \cite{balk} gluon exchange with the heavy quark at
the transition vertex  can give  additional momentum to the light quark (light
degrees of freedom) after the transition, (the so--called velocity--kick
 \cite{koerner}), which goes beyond the approximation of the present paper.

\section{CONCLUSIONS}

The authors of ref.\cite{close} showed how to match quark model
and HQET in the
particular case of electromagnetic transitions and  where all interactions
between the constituents were ignored. In \cite{close} a non--trivial
structure emerged as a result of dealing carefully with the Wigner--rotation.
The essential features generalized to the case of present interest and enabled
a compact description of mesons suitable for EHQT (eq\ref{meson2}).

We applied these ideas to an explicit
calculation of the form factors for heavy flavour transitions
involving S-wave states. We found that the calculations of
the non-relativistic ISGW model (ref \cite{scora})
 are changed significantly and that
the effect of Wigner rotations merits care.

In particular
we found that the model parameters required to fit the spectroscopy
\cite{scora} lead to an excellent description of the slope of the
universal form factor, in contrast to previous
literature. Specifically, in
this quark model calculation the $\rho$ parameter which is defined
 via $\xi(y) = 1 - \rho^2 (y - 1) + {\rm O}((y-1)^2) $ turned out to be:
\begin{center}  $\rho = 1.19 \pm 0.02$ \end{center}
This describes an enhancement of the effect that the light degrees of freedom
have
compared with the simple quark model calculation which gave $\rho = 0.568$ and
 the relativistic calculation without Wigner rotation: $\rho =1.03$.
This result flowed from the expansion of the current
operator to $O(\vec{v}^2)$ as required for consistency.

It is interesting to note the parallel with similar work
in the light quark sector
where Close and Li \cite{li} pointed out that
the apparent difference between parameters for fitting
spectroscopy \cite{karl} and current induced transitions
($\gamma$ and electroproduction) were artefacts of an inconsistent
non--relativistic restriction. When calculations of current transitions were
made consistently to $O(\vec{v^2})$ both dynamics and spectroscopy could be
simultaneously described \cite{li}.

Our work highlights the role of the two independent ($\vec{k},\vec{v}$)
boosts. On physical grounds we expect that these considerations are more
general
than the specific model. However, in turn, the present approach has its
limitations.
We have shown that the expectation of current operators between spinors of
massive quarks in composite systems leads to significant deviations from
naive  NRQM but, as shown in ref \cite{li}, the binding potential can also
make an explicit contribution to the current operator. The major effect from
a scalar potential is to renormalise the quark mass; insofar as this is a
parameter in the model, it is effectively incorporated in the present work.

Our calculations provide a  measure of the validity
 of Heavy Quark Effective Theory in finite mass situations.
The Wigner spin rotations arise when boosting a state whose constituents
$\frac{\vec{k}}{m} \neq 0$ in the overall rest frame. For b-quarks this effect
is less than 10\% and can be ignored: HQET applies. For c-quarks the effect
of $(\frac{\vec{k}^2}{Mm})$ is of order up to 30\% and so caution should be
exercised at this level. For strange quarks this term is of the
 order 70\% and can not be ignored. This deviation from the infinite
mass limit shows itself e.g. in the mixing of
the $^{3}P_1$ and $^{1}P_1$ states in the Kaon--system, where in the
heavy mass limit one would expect a mixing angle of
 $35^0$ \cite{koerner}, while experiments indicate
an angle of about $45^0$ \cite{XXX}.
This may modify some of the analysis of $B \rightarrow K^{**} \gamma$
\cite{ali}. This entails extension of these ideas to P-states and is
under investigation \cite{achim}.

\vskip 0.3in

We are indebted to A.Le Yaouanc and J.C.Raynal for helpful comments. A.W.
thanks
the German Academic Exchange Service and the German Scholarship Foundation for
financial support.


\begin{thebibliography}{20}

\bibitem{isgur1} N.\ Isgur\ and M.B.\ Wise, Phys.\ Lett.\ B 237 (1990) 527
\bibitem{voloshin} M.\ Voloshin, Phys.\ Rev.\ D 46 (1992) 3062
\bibitem{ali} A.\ Ali, T.\ Ohl and T.\ Mannel, Phys.\ Lett.\ B298 (1993) 195
\bibitem{neubert2} M.\ Neubert and V. Rieckert, Nucl.\ Phys.\ B 382 (1992) 97
\bibitem{ukqcd} UKQCD--collaboration, private communication and
preprint Edinburgh 93/525; Southampton SHEP 92/93-17
\bibitem{blok} B.\ Blok and M.\ Shifman, \ Phys Rev D47 (1993) 2949
\bibitem{rad}   A.\ V.\ Radyushkin, Phys.\ Lett.\ B271 (1991) 218
\bibitem{neubert3} M.\ Neubert, Phys.\ Rev.\ D 45 (1992) 2451
\bibitem{scora} N.\ Isgur, D.\ Scora, B.\ Grinstein, M.\ Wise, Phys.\
Rev.\  D 39 (1989) 799
\bibitem{neubert1} M.\ Neubert, Phys.\ Lett.\  B 264 (1991) 455
\bibitem{rosner} J.\ L.\ Rosner, Phys.\ Rev.\  D 42 (1990) 3732
\bibitem{mannel} T.\ Mannel, W.\ Roberts, Z.\ Ryzak, Phys.\ Lett.\ B
254 (1991) 274\\
                T.\ Mannel, IKDA 92/6
\bibitem{li} F.\ E.\ Close and Z.\ Li, Phys.\ Rev.\ D 42 (1990) 2194\\
             F.\ E.\ Close and Z.\ Li, Phys.\ Rev.\ D 42 (1990) 2207
\bibitem{close} F.\ E.\ Close and Z.\ Li, Phys.\ Lett.\ B237 (1992) 143
\bibitem{wirbel} M.Wirbel, B.Stech and M.Bauer, Z.Phys C29, (1985) 637
\bibitem{balk} S.\ Balk, J.\ G.\ K{\"o}rner, G.\ Thompson, F.\
Hussain, IC-91-397, June 1992
\bibitem{falk} A.\ F.\ Falk, Nucl.Phys. B 378 (1992) 79
\bibitem{bjorken} J.\ D.\ Bjorken, invited talk given at ``Les Recontre de la
Valle d'Aosta La Thuile'' Editions Frontiers, Gif-Sur-Yvette:(1990) 583
\bibitem{brodsky} S.\ Brodsky and J.\ Primack, Ann.\ Phy.\ 52, 315 (1969)
\bibitem{iddir} E.\ Golowich et al., Phys. Lett. B 213 (1988) 521
\bibitem{achim} A.Wambach, in preparation
\bibitem{pene} A.LeYaouanc et al., Ann.\ Phy.\ 88, 242 (1974) and Erratum 97,
                  567 (1976)\\
               A.LeYaouanc et al.,Phys.\ Rev.\ D 15 (1977) 844
\bibitem{koerner} J.\ G.\ K{\"o}rner, Nucl.Phys. B (Proc. Suppl.) 21 (1991) 366
\bibitem{karl} N.\ Isgur and G.\ Karl, Phys.\ Rev.\ D 18 (1979) 4187\\
               N.\ Isgur and G.\ Karl, Phys.\ Rev.\ D 19 (1979) 2653
\bibitem{XXX} G.\ W.\ Brandenburg et. al., Phys.\ Rev.\ Lett.\ 36 (1976) 703\\
              R.\ K.\ Carnegie et. al., Phys. Lett. B 68 (1977) 287




\end{thebibliography}
\end{document}